\documentstyle[11pt]{article}

\def\al{\alpha}
\def\d{\partial}

\def\te{\vartheta}

\begin{document}

\title{Is dark matter a phantom ? }
\author{G\"unter Scharf \footnote{e-mail: scharf@physik.unizh.ch}
\\ Institut f\"ur Theoretische Physik, 
\\ Universit\"at Z\"urich, 
\\ Winterthurerstr. 190 , CH-8057 Z\"urich, Switzerland}

\date{}

\maketitle\vskip 3cm

\begin{abstract} We show that Einstein's equations in a non-standard gauge have vacuum solutions with an asymptotically flat rotation curve as it is observed in the dark halos of galaxies. Introducing a material disk into this model we find a matter density in accordance with the Tully-Fisher relation.

\end{abstract}
\vskip 1cm
{\bf PACS numbers: 04.06 - m; 04.09 + e}

\newpage

Choosing spherical coordinates $(t, r, \te, \phi)$ we write the static, spherically symmetric line element in the form
$$ds^2=g_{\al\beta}dx^\al dx^\beta=e^adt^2-e^bdr^2-r^2e^c(d\te^2+\sin^2\te d\phi^2).\eqno(1)$$
where the metric functions $a(r), b(r), c(r)$ depend on $r$ only, and the speed of light is $=1$. In (1) one usually sets $c=0$ by redefining the radial coordinate $r$. This is the standard gauge which is assumed in the proof of Birkhoff's theorem and leads to the Schwarzschild metric.

Using this gauge to describe galaxies one is then forced to postulate dark matter in order to get the observed flat rotation curves right. However, one may have doubts whether this is the right idea because physically $r$ is defined by the measuring process of the astronomer and with the above redefinition one loses contact to physics. To be on the safe side we keep $c(r)\ne 0$, but then, clearly, the field equations do not determine all three functions $a, b, c$ uniquely. To choose a physical gauge fixing condition we consider the circular velocity $V(r)$. From the geodesic equation on finds the following expression for $V(r)$ ([2], eq.(2.14))
$$\vec V^2={g_{\phi\phi}\over g_{tt}}{\Gamma_{tt}^r\over\Gamma_{\phi\phi}^r},\eqno(2)$$
where the $\Gamma$'s are the Christoffel symbols for the metric (1).
We now take this relation as our non-standard gauge fixing condition [1]. That means $V(r)$ must be given, it cannot be predicted from the vacuum equations alone. As we shall see it is the other way around: $V(r)$ determines the metric (1) uniquely.

The non-vanishing components of the Ricci tensor for the metric (1) are the diagonal elements [1]
$$R_{tt}={1\over 2} e^{a-b}(a''+{1\over 2} a'^2-{1\over 2} a'b'+a'c'+{2\over r}a')\eqno(3)$$
$$R_{rr}=-{1\over 2}(a''+2c'')+{b'\over 4}(a'+2c'+{4\over r})-{a'^2\over 4}-{c'^2\over 2}
-{2\over r}c'\eqno(4)$$
$$R_{\te\te}=e^{c-b}[-1-{r^2\over 2}c''-r(2c'+{a'-b'\over2})-{r^2\over4}c'(a'-b'+2c')]+1
\eqno(5)$$
$$R_{\phi\phi}=\sin^2\te R_{\te\te},\eqno(6)$$
the prime always denotes $\d /\d r$. Then the Einstein's equations without matter can be reduced to the following three differential equations  
$$G_{tt}=e^{a-b}\Bigl[-c''-{3\over 4}c'^2+{1\over 2}b'c'+{1\over r}(b'-3c')\Bigl]+{1\over r^2}(e^{a-c}-e^{a-b})=0\eqno(7)$$
$$G_{rr}={1\over 2}a'c'+{1\over r}(a'+c')+{c'^2\over 4}+{1\over r^2}\Bigl(1-e^{b-c}\Bigl)=0\eqno(8)$$
$$G_{\te\te}={r^2\over 2}e^{c-b}\Bigl[a''+c''-{1\over r}(b'-a'-2c')+{1\over 2}(a'^2-a'b'+a'c'-b'c'+c'^2)=0.\eqno(9)$$

It is not hard to see [2] that there are only two independent field equations. Indeed, using (8) $b$ can be expressed by $a$ and $c$. Eliminating $b$ there remains one second order differential equation for $a$ and $c$:
$$c''={a''\over a'}\Bigl(c'+{2\over r}\Bigl)+{4\over r^2}+a'c'+{c'^2\over 2}+{2\over r}(a'+c').\eqno(10)$$
Introducing the new metric function
$$f(r)=c(r)+2\log {r\over r_c}\eqno(11)$$
where $r_c$ has been included for dimensional reasons, equation (10) assumes the simple form
$${f''\over f'}-{a''\over a'}=a'+{f'\over 2}.\eqno(12)$$
This can immediately by integrated
$$\log{f'\over a'}=a+{f\over 2}+{\rm const.}\eqno(13)$$
On the other hand the circular velocity squared (2) expressed by the metric functions simply becomes
$$V^2(r)\equiv u={a'\over f'}.\eqno(14)$$
Using this in (13) we have
$$a+{c\over 2}+\log{r\over r_c}=-\log u.\eqno(15)$$
Differentiating and eliminating $c'$ by means of (11) and (14)
$$c'=f'-{2\over r}={a'\over u}-{2\over r},\eqno(16)$$
we find
$$a'=-{2u'\over 1+2u}.\eqno(17)$$
This gives the first diagonal element of the metric
$$g_{tt}=e^a={K_a\over 1+2u}\eqno(18)$$
where $K_a$ is a constant of integration. Then from (15) we  get
$$g_{\te\te}=e^c=\Bigl({1+2u\over u}\Bigl)^2{K_c\over r^2},\eqno(19)$$
where $K_c$ is another integration constant which contains $r_c$. Finally, $\exp b$ follows from (8)
$$g_{rr}=e^b=K_c\Bigl({u'\over u^2}\Bigl)^2(1+2u).\eqno(20)$$
We have succeeded in expressing the metric by the circular velocity squared $u(r)$. If we choose $u=r_s/2(r-r_s)$, we recover the standard gauge $c=0$ and the Schwarzschild metric. But now also flat rotation curves are possible without dark matter. {\it As the ether is superfluous in special relativity, so seems to be dark matter in general relativity.} From (18-20) we are able to predict other observable quantities which can be computed from the metric, for example lensing data [3]. In this way the theory can be tested.

Another test is to introduce normal matter located in the equatorial plane which simulates a spiral galaxy. There exists a simple method to construct solutions of Einstein's equations with such a disk. This is the so-called displace, cut, and reflect method which goes back to Kuzmin [4] and since then was used and modified by many authors (see [5] and references given there). If the equatorial plane is $z=0$ in cylindrical coordinates, the matter density is singular $\sim\delta (z)$ and the metric $g_{\mu\nu}$ has jumps in its first derivatives. The mathematical basis in this situation is Taub's theory of distribution valued curvature tensors [6]. The Einstein tensor then has a singular part which is related to the distribution valued energy-momentum tensor by
$$(R_{\mu\nu}-{1\over 2}g_{\mu\nu}R)\vert_{\rm sing}=\kappa t_{\mu\nu}\delta(z),\eqno(21)$$
where
$$R=g^{\al\beta}R_{\al\beta},\quad \kappa={8\pi G\over c^2}.$$
If the jumps $b_{\mu\nu}$ of the normal derivatives of $g_{\mu\nu}$ are defined by
$$[g_{\mu\nu,\sigma}]\equiv{\d g_{\mu\nu}\over\d x^\sigma}\Bigl\vert_+-{\d g_{\mu\nu}\over\d x^\sigma}\Bigl\vert_-
=n_\sigma b_{\mu\nu},\eqno(22)$$
where $+$ and $-$ mean the limiting values from both sides of the singular plane with normal vector $n_\sigma$, then
$t_{\mu\nu}$ can be calculated from ({6], eq.(6-2))):
$$-2\kappa t_{\mu\nu}=n^2\Bigl((g^\sigma_\mu-{n^\sigma n_\mu\over n^2})(g^\tau_\nu-{n^\tau n_\nu\over n^2})-$$
$$-(g_{\mu\nu}-{n_\mu n_\nu\over n^2})(g^{\sigma\tau}-{n^\sigma n^\tau\over n^2})\Bigl)b_{\sigma\tau},\eqno(23)$$
where $n^2=n^\al n_\al$.

We go over to cylindrical coordinates $(t,R,z,\phi)$
$$r^2=R^2+z^2,\quad z=r\cos\te,\quad \sin\te={R\over r}.$$
Then the metric (1) assumes the following non-diagonal form
$$ds^2=g_{\mu\nu}dx^\mu dx^\nu$$
with
$$g_{00}=e^a,\quad g_{11}=-{R^2\over r^2}e^b-{z^2\over r^2}e^c,\quad g_{22}=-{R^2\over r^2}e^c-{z^2\over r^2}e^b$$
$$g_{12}=g_{21}=-2{z\over r}(e^b-e^c),\quad g_{33}=-R^2e^c.\eqno(24)$$
For simplicity we still use $r$, but our admissible coordinates are $x^1=R, x^2=z$.

We now apply the displace, cut, and reflect method following the procedure of Voigt and Letelier [5]. We take the metric (24) with $a, b, c$ given by (18-20) in the half space $z>d>0$, displace it to $z=0$ and reflect it for $z<0$. This produces the finite jumps in the $z$-derivatives of $g_{\mu\nu}$. The whole procedure is equivalent to the transformation $z\to |z|+d$. The normal vector is $n_\mu=(0,0,1,0)=\delta^2_\mu$ and
$$n^\nu=g^{\nu\mu}n_\mu=g^{\nu 2},\quad n^\nu n_\nu=g^{22}.$$
The jumps (22) in the normal derivatives on $z=0$ which we need are equal to
$$b_{11}=[g_{11, 2}]=g'_{11}{2d\over r}-{4d\over r^2}e^c\eqno(25)$$
$$b_{33}=[g_{33, 2}]=g'_{33}{2d\over r},$$
where the prime always means $\d /\d r$ keeping $z$ and $R$ constant. Now from (23) we find the energy density
$$t_0^0={1\over 2\kappa}\Bigl(Db_{11}+{g_{11}\over Dg_{33}}b_{33}\Bigl)\eqno(26)$$
with $D=g_{11}g_{22}-g_{12}^2$. Using
$$b_{11}={2d\over r}\Bigl(g'_{11}-{2\over r}e^c\Bigl),\quad b_{33}={2d\over r}g_{33}c',$$
we finally obtain
$$t_0^0=-{d\over\kappa r}\Bigl(e^{b+c}{R^6\over r^4}\d_r({e^b\over r^2})+{2R^4\over r^5}e^{b+2c}-{r^2\over R^2}\d_re^{-c}\Bigl).\eqno(27)$$
Here we have to put $z=0$ everywhere which gives $r^2=R^2+d^2$. 

 Now we must specify the circular velocity squared $u(r)$ in order to fix the metric. We are particularly interested in the case of an asymptotically flat circular velocity which in the usual terminology corresponds to a dark halo. Therefore we assume $u(r)$ of the form
$$u(r)=u_{\rm flat}+{u_1\over r}+O(r^{-2})\eqno(28)$$
for large $r$.  Then it follows from (18-20)
$$e^a=K_a+O(r^{-1}),\quad e^b={L_b\over r^4}+O(r^{-5})\eqno(29)$$
$$e^c={L_c\over r^2}+O(r^{-3})\eqno(30)$$
 where by (19)
$$L_c\sim u_{\rm flat}^{-2}=V_{\rm flat}^{-4}.\eqno(31)$$
 Using this in (27) the leading order comes from the last term
$$t_0^0={2d\over\kappa L_c}{r^2\over R^2}(1+O(R^{-1})).\eqno(32)$$
This is proportional to the density of normal matter because we consider a static energy-momentum tensor. Taking (31) into account we find that
$$t_0^0\sim u^2_{\rm flat}\sim V^4_{\rm flat}(R)\eqno(33)$$
for large $R$. This is in accordance with the baryonic Tully-Fisher relation for galaxies [7] [8], which states that the total baryonic mass $M$ is proportional to $V^4_{\rm flat}$. In fact, one can show [9] that the contribution of the inner part $R<R_1$ of the disk can be made arbitrarily small compared to the outer part between $R_1<R<R_2$, say. We emphasize that $M$ is obtained from $t_0^0$ by integrating with the Euclidean surface measure $R\, dR\, d\phi$, because this is what astronomers are doing when they determine $M$ from luminosity measurements.
  
The radial pressure $t^r_r$ vanishes because $G_rr$ (8) does not contain a second derivative. Therefore our model must be interpreted as a dust disk with purely azimuthal stresses. This is not very realistic and it remains to be investigated whether the Tully-Fisher relation is a generic property for more physical galaxy models. But at present the answer to the question in the title is a careful ``may-be yes''.

\end{document}